\documentclass[article,number,sort&compress]{elsarticle} 

\biboptions{sort&compress,square}
\usepackage{lineno}

\newcommand{\fem}{f_{em}}
\newcommand{\vev}[1]{\left\langle#1\right\rangle}

 \usepackage{graphicx}
 
\usepackage{amssymb}

 \usepackage{color}

\begin{document}

\journal{Nuclear Instruments and Methods}

\begin{frontmatter}

\title{Degradation of resolution in a homogeneous dual-readout hadronic calorimeter}


\author{Donald E. Groom}

\address{Lawrence Berkeley National Laboratory,
50R6008, Berkeley, CA 94720, USA}
\corref{Tel: 1 510 486 6788; fax: +1 510 486 4799}
\ead{degroom@lbl.gov}

\begin{abstract}
If the scintillator  response to a hadronic shower in a semi-infinite uniform calorimeter structure is 
$S$ relative to the electronic response, then $S/E = [\fem + (1-\fem)(h/e)]$, where $E$ 
is the incident hadron energy, $\fem$ is the electronic shower fraction, and $h/e$ is the 
hadron/electron response ratio.  If there is also a simultaneous readout with a different $h/e$, say a Cherenkov 
signal $C$, then a linear combination of the two signals provides an estimator of $E$ that is
proportional to the incident energy and whose distribution is nearly Gaussian---even though
the $S$ and $C$ distributions are non-linear in $E$, wide, and skewed.
Since an estimator of $\fem$ is also obtained, it is no longer a stochastic variable. Much of the remaining
resolution variance is due to sampling fluctuations. These can be avoided in a homogeneous 
calorimeter. The energy resolution depends upon the contrast in $h/e$ between the two channels.
$h/e$ is small in the Cherenkov channel. \it Mechanisms that increase   $h/e$ 
in sampling  calorimeters with organic  scintillator readout are not available in a homogeneous inorganic scintillator calorimeter\rm. The $h/e$ contrast is very likely too small to provide the needed energy resolution.  
\end{abstract}


\end{frontmatter}


\section{Introduction}

A homogeneous dual-readout hadron calorimeter has been suggested for possible use at a future 
linear collider~\cite{para10}.  The machine will probably be an $e^+e^-$ collider and in some concepts
will have a long bunch spacing (${\cal{O}}$(100\,ns)), so that detectors with  time constants in this 
range might be used.  
Discrimination between the Cherenkov signal ($C$) and scintillator ($S$) optical signals is expected to use a combination of timing, color, and, possibly, Cherenkov light direction and polarization.

In practice corrections must be made for cracks, leakage, and light collection variations, 
and the structure usually varies with depth.  For the purposes of this analysis 
we shall assume that the corrections have been made properly, 
and consider a semi-infinite calorimeter with uniform structure that is either fine-sampling or homogeneous.

In each high-energy interaction in a hadronic cascade an average of 1/4 of the energy 
is carried away by $\pi^0$'s~\cite{gabriel94}.  These immediately decay to $\gamma$'s 
which initiate electromagnetic  (EM) showers.   This occurs many times, with the result that a large fraction 
$\fem$ of the incident energy joins the EM shower.  The mean, $\vev{\fem}$, is
$\approx 0.5$ for 100--150~GeV incident pions.  
It increases slowly with incident energy $E$, asymptotically approaching unity. 

The hadronic response $S$ to an incident hadron with energy $E$ (calibrated to electron response) is
\begin{equation}
S = E  [\,\fem+(1-\fem)   (h/e) ] \ .
\label{eqn:hadresponse}
\end{equation} %
 The EM energy deposit is detected with relative efficiency $e$, and the hadronic 
signal with relative efficiency $h$. Both vary from event to event. In part because of low multiplicities in the initial hadronic interactions, the variance of $h$ is much
 larger than the variance of $e$. It makes sense to treat $h/e$ as a stochastic variable.   To the extent 
 that the variance of $h$ dominates, the distribution of the conventional $e/h$
 is not useful.  In Sec.~\ref{sec:resolutionsection} we treat the distribution of $h/e$ as Gaussian. 
 
 Most energy deposit is by very low-energy electrons and charged hadrons.  Because so many 
 generations are involved in a high-energy cascade, the
 hadron spectra are essentially independent of the cascade's origin except for overall 
normalization. This ``universal spectrum'' concept is discussed in detail in 
 Ref.~\cite{gabriel94}.  It is because of this feature that $\vev{h/e}$ is a robust quantity, 
 independent of energy and  incident hadron species.
 
The energy-independent $\vev{h/e}$ does 
depend upon calorimeter composition and structure, as well as the 
readout---for example, an organic scintillator readout is sensitive to the otherwise-invisible 
neutron content of the cascade while a Cherenkov readout is relatively 
blind to the hadronic content. $\vev{\fem}$ can be found by fitting the average $\pi^-/e$  
response as a function of test-beam energy with an appropriate 
$\vev{\fem}$ parameterization such as a power law in 
energy~\cite{gabriel94}.\footnote{Technically, a power-law fit finds  $a = (1-\vev{h/e})E_0^{1-m}$.
Since $1-m$ is small and the scale energy $E_0$ is close to 1~GeV for pion-induced cascades, 
the distinction is minor: $\vev{h/e}\approx 1-a$. A similar  distinction occurs when other 
parameterizations are used.  $\vev{h/e}$ itself cannot be isolated.}

Usually $\vev{h/e}$ is less than unity,  since the EM contribution is detected 
with greater efficiency than the hadronic energy deposition.
If $\vev{h/e}$ is not unity, then the broad, skewed $\fem$ probability distribution function 
(p.d.f.) significantly degrades and skews the
energy resolution, resulting in  the familiar wide, non-Gaussian energy 
distributions. The response is not linear with energy because of the energy dependence of $\vev{\fem}$.
 If $\fem$ could  be \it measured\rm\ for each event, then the 
response as given in Eq.~\ref{eqn:hadresponse} could be corrected to the actual energy with a 
nearly Gaussian distribution and a mean proportional to the energy.

The importance of measuring the EM content on an event-by-event basis was realized as early as 1980, 
although how to use the information was not so clear.  There was even a (stillborn) 
dual readout test by A. Erwin (BNL) using scintillator
and radiator plates~\cite{hauptman}.  

EM showers result in large
local energy deposit; with sufficient readout segmentation this ``lumpiness'' provides a
measure of $\fem$.  Weighting this part differently than the remaining signal might improve resolution. 
This approach was used with some success by the WA1
collaboration~\cite{WA1_1981}, but has been less successful elsewhere, 
e.g., for the \hbox{ATLAS} central barrel calorimeter~\cite{ATLAS_weighting}.  

In a 1983 summer school review of high-energy calorimetry, P. Mockett stressed the 
importance of measuring the fractional EM content of the shower. He speculated that one could use two sampling media, an electron-sensitive detector (Cherenkov) and an ionization sensitive detector 
(scintillator). He also imagined taking advantage of the fast Cherenkov pulse and slow 
scintillation signal in a heavy 
inorganic scintillator. Both suggestions were prophetic~\cite{mockett83}. 

Such a separation was actually made by Theodosiou et al.~\cite{theodosiou84} in 1984, using the time structure of pulses observed in scintillating glass.  He thought the technique might permit electron/hadron separation or even help with particle identification.  Winn later suggested using color in addition to timing to make the 
separation~\cite{winn89}.  There must have been considerable speculation about dual-readout calorimetry,
but only Theodosiou et al.\ took this speculation into the laboratory.

Part of the problem was that the physics of energy deposition had not yet been elucidated, or at least widely
understood. This came in the late 1980s  with the work of Fabjan et al.~\cite{Ufission_Fab77}, Wigmans~\cite{wigmans87}, Br\"uckmann et al.~\cite{bruckmann88}, Drews et al.~\cite{drews90},
 and others, but a key element was the energy deposition inventories produced by the very 
 detailed simulations of Gabriel and his collaborators at Oak Ridge as early as 1974~\cite{gabriel74,Ufission_Fab77}.  
 
 Much of the hadronic energy resolution problem was related
to the large fraction of missing energy in the hadronic sector, due to nuclear dissociation, nuclear recoil,
residual nuclear excitation, $\mu$ and $\nu$ escape, and (unobserved) neutrons.  Scintillator response
to highly ionizing charged particles is non-linear, resulting in significantly more lost signal.  For a time it was 
thought that ionization by U fission products could make up some of the lost 
energy~\cite{Ufission_Fab77,leroy86,wigmans87}, but non-linear scintillator response to the 
highly ionizing fragments negated most of the gains.

In a sampling calorimeter, only a small fraction of the energy is deposited in the sensors (quartz or scintillator), and fluctuations in this fraction are more important than intrinsic fluctuations in the hadronic signal.  These
dominate, once $\fem$ is removed.  The sampling fluctuations are avoided in a  homogeneous
 calorimeter.  The possibly long bunch spacing at a future linear
collider opens the door to a homogeneous dual-readout dense crystal or glass calorimeter, 
where a fast, blue, Cherenkov pulse might be separated from a slower, redder scintillation signal. 
Crystal studies are being successfully explored by 
Akchurin, et al.~\cite{Ak05,Ak08,Ak08bb,Ak09,Ak09bb,Ak09cc,dopedPbWO4,Ak11}, but with only speculative 
mention of dual-readout hadron calorimeters.  
A feasibility study is part of a new proposal~\cite{superDREAM2010}.

Akchurin et al.\ have demonstrated  signal separation that would be adequate for recovering energy 
estimators that are linear in the corrected energy and have a nearly Gaussian distribution.  However, I am 
concerned that the energy \it resolution\rm\ would not be adequate.
In this paper I explore the likely resolution as a function of energy 
and $\vev{h/e|_S}$ using resolution contributions based on published crystal, glass, 
and sampling calorimeter performance.  
Simple, transparent Monte Carlo simulations (MC's) are used by choice, 
to make the physics more transparent than if a sophisticated MC such as GEANT4 were used.  
The p.d.f.\ of $\fem$ is approximated with some care,
while other resolution contributions are taken to be Gaussian.


\section{$\vev{h/e}$ in a high-density crystal or glass scintillator}\label{hoe_glass}

In an EM cascade the electrons are relativistic until their energies fall well 
below the critical energy, so that almost all of the energy is deposited by near-minimum 
ionizing electrons. No appreciable energy exits from the EM cascade via   
photonuclear interactions. 
The result is a response very nearly linear in the incident electron or photon energy.  

Hadronic interactions deposit energy in a variety   of ways.  (An inventory is given in Table~1 [by Gabriel and Schmidt] in Ref.~\cite{Ufission_Fab77}, and detailed discussions can be found in 
Refs.~\cite{wigmansbook,leroy_rancoita,ak_wig_2012_review}  and other recent reviews).
A large fraction of the hadronic energy ($\approx20\%$ for Fe/scintillator and 
$\approx 40\%$ for U/scintillator sampling calorimeters) goes to nuclear 
dissociation and recoil, and is ``invisible.''     Neutrinos and most muons escape.  
Some fraction of the neutrons 
can be detected via $n$--$p$ scattering in hydrogeneous materials such as organic scintillator, 
but much or most of the neutron energy is also lost.   Low-energy protons and charged 
fission fragments produce saturated signals in scintillator.   (This
occurs in inorganic~\cite{inorganic_saturation} as well as organic scintillators~\cite{BirksLaw}.) 
All of these factors result in low visible 
response to the hadronic component of the cascade relative to response to the EM component.

Detection of recoil protons in neutron scattering in hydrogenous detectors increases 
$h$~\cite{Gal86}.  In a sampling calorimeter a  disproportionate fraction of the EM energy 
is deposited in the higher-$Z$ absorber;
the absorber/active region thickness ratio can be ``tuned'' to decrease   $e$. 
 Both of these effects increase $h/e$. In practical sampling calorimeters $\vev{h/e}$ 
 is typically 0.7, and can be made to approach unity with careful design.
\vskip1mm
\it Neither mechanism for increasing $h/e$ is available to a high-density 
homogenous calorimeter.\rm
\vskip1mm
As we shall see, the resolution is dependent on the ``$h/e$ contrast,'' the difference between 
$\vev{h/e}$ for the Cherenkov ($\vev{h/e|_C}$) and scintillation($\vev{h/e|_S}$) readouts.  
Based on experience with quartz-fiber readout 
calorimeters~\cite{qcal97,Wigmans_Perugia_04,DREAM05}, $\vev{h/e|_C}=0.20$--0.25.%
\footnote{From the data
shown in Table~3 of Ref.~\cite{qcal97} I obtain $h/e|_{_C}=0.247$~\cite{groom07}.}
 There are few data concerning $\vev{h/e|_S}$ in a homogeneous calorimeter, but there 
is no way to hide EM energy in the absorber and there is very little neutron sensitivity.   
We might expect as much as 30\% of the hadron energy to be expended on 
nuclear dissociation and therefore invisible, and 
15\%--20\% to be carried by neutrons.  These alone would result in  
$\vev{h/e|_S} \approx 0.5$.  There
are other effects, such as incomplete Cherenkov-scintillator separation and saturated 
scintillator response to highly ionizing particles, 
so $\vev{h/e|_S} = 0.35$--0.5 might be expected.  This is corroborated by a comment in 
Ref.~\cite{Ak08}: ``The $e/h$ value of ECAL [PbWO$_4$] as a scintillation device is much larger than for the Cu/plastic sampling structure in DREAM: 2.4 vs.~1.3.''  ($h/e=0.43$ vs 0.7.)


\section{Dual-readout hadronic calorimetry}\label{dualcal}

\begin{figure}
\centerline{\includegraphics[height=3in]{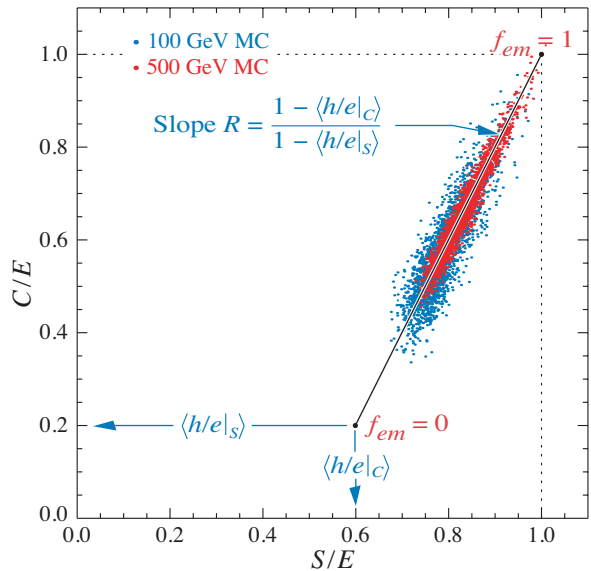}}
\caption{Energy-independent event locus in the $C/E$--$S/E$ plane. As indicated by the MC events, resolution improves and the mean moves upward along the locus as the beam energy is increased.}
\label{fig:locus}
\end{figure}

In 1997 Wigmans discussed the advantages of adding a quartz-fiber readout to the scintillator
readout of a sampling calorimeter~\cite{tucson97}.  
With a Cherenkov readout ($C$) fairly blind to 
hadronic activity and a scintillator readout ($S$) with optimized hadronic response, $\fem$ could be
determined (or eliminated) and a corrected energy found.
The DREAM collaboration elegantly  implemented this proposal 
with a test-beam calorimeter having quartz  and plastic scintillator fibers in copper 
tubes~\cite{Wigmans_Perugia_04,DREAM05,DREAM07}.

In the dual-readout case Eq.~\ref{eqn:hadresponse} is replaced 
by~\cite{Wigmans_Perugia_04,groom07,RPP12,groom13}
\begin{eqnarray} 
S&=& E[  f_{em}+ (1- f_{em})   (h/e|_S) ]
\label{eqn:S0}\\
C &=&  E[ f_{em} + (1-f_{em})   (h/e_C)]   \  .
\label{eqn:C0}
\end{eqnarray}

In parametric form, Eqns.~(\ref{eqn:S0}) and~(\ref{eqn:C0})
describe a straight line-segment event locus in the $C$--$S$ (or $C/E\hbox{--}S/E)$ 
plane, as illustrated in Fig.~\ref{fig:locus}.
If the cascade is ``all electomagnetic'' ($\fem= 1$), then $S/E = C/E = 1$.
If the cascade is ``all hadronic'' ($\fem= 0$), then $S/E = h/e|_S$ and $C/E = h/e|_C$.  
(The MC ``events'' shown in the figure are discussed below.)  The slope in the 
$C/E\hbox{--}S/E$ plane is independent of energy; with increasing energy the distribution
just moves up along the locus.

It is convenient to introduce  the less cumbersome notation $h/e|_X \equiv \eta_X$:
\begin{eqnarray} 
S &=& E [f_{em}+ (1- f_{em})  \, \eta_S]
\label{eqn:S}\\ 
C &=&  E[ f_{em} + (1-f_{em}) \, \eta_C] 
\label{eqn:C}
\end{eqnarray}

Equations~(\ref{eqn:S}) and~(\ref{eqn:C}), linear in $1/E$ and $\fem$,  can be rewritten as
\begin{equation}
\left( \matrix{ S &-(1- \eta_S)\cr C& -(1- \eta_C) }\right)
 \left( \matrix{1/E \cr \fem }\right)  
= \left( \matrix{\eta_S \cr \eta_C }\right) 
\label{eqn:matrixversion}
\end{equation}
with solutions~\cite{groom13}
\begin{eqnarray}
E &=& \displaystyle \frac {S(1- \eta_C) - C(1- \eta_S)}{\eta_S -  \eta_C}
\label{eqn:Elongform}  \\
\fem &=& \frac      {C\,\eta_S-S\,\eta_C}     {S\,(1-\eta_C) -C(1-   \eta_S)  }
\label{femSolution}\ .
\end{eqnarray}

There is an important difference between Eqs.~\ref{eqn:hadresponse}, \ref{eqn:S0}, and \ref{eqn:S},
and Eq.~\ref{eqn:Elongform}:  The first three give estimators of the scintillator response, 
given $\fem, \  \eta_S$, and the incident energy $E$.  In contrast, Eq.~\ref{eqn:Elongform} provides an 
\it  estimator\rm\ of this energy given $S$, $C$, $\eta_S$ and $\eta_C$.  Similarly, Eq.~\ref{femSolution}
provides an  \it estimator\rm\ of $\fem$.

The sensitivity of the energy estimator to the $h/e$ contrast is particularly manifest in Eq.~\ref{eqn:Elongform}:
If $\vev{\eta_S}-\vev{\eta_C}$ is small compared to the statistical fluctuations of $\eta_S$ and $\eta_C$,
then the scatter in the energy estimators will be large.

In Eqs.~\ref{eqn:Elongform} and \ref{femSolution},  $\eta_C$ an $\eta_S$ are the 
values \it peculiar to that event.\rm\
These are unknown---and unknowable, until some way of tagging the hadronic composition 
becomes available.\footnote{Neutron detection has been proposed 
and is being explored for this purpose~\cite{Ak07_neutrons,Ak09_neutrons}.}
But in an experimental situation, an estimator of the energy must be established for each
event. There is little choice but to replace these quantities by their means.  

In this case, it is convenient to write the energy estimator  (Eq.~\ref{eqn:Elongform}) more
compactly as~\cite{groom07,RPP12,groom13}
\begin{equation}
E = \frac{RS-C}{R-1} \ ,
\label{eqn:Eshortform}
\end{equation}
where I have made use of the slope of the event locus (the ratio of ranges of $S$ and $C$) 
shown in Fig.~\ref{fig:locus}:
\begin{equation}
R \equiv \displaystyle \frac {1- \vev{\eta_C}}{1- \vev{\eta_S }}
\label{eqn:slopedefinition}
\end{equation}


\section{Electromagnetic fraction}\label{femsection}

In a dual-readout calorimeter estimators of $E$ and $\fem$ can be determined on 
an event-by-event basis.  The EM fraction $\fem$ is  no longer a stochastic variable.

\begin{figure}
\centerline{\includegraphics[width=3in]{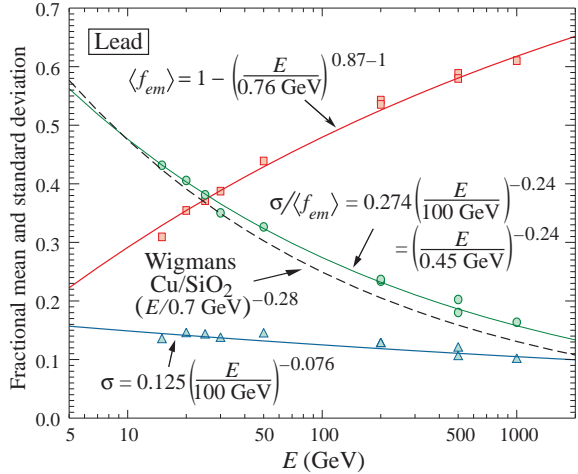}}
\caption{The mean, standard deviation, and fractional standard deviation of $\fem$ in lead,
as simulated by FLUKA90.  This is basically  Fig.~3(b) in Ref.~\cite{groom07}. An error
in the fit to the fractional standard deviation has been corrected,
a slightly different MC data set has been used, and Wigmans' fit to the 
fractional standard deviation in a copper/quartz-fiber calorimeter has been 
added~\cite{wigmansbook}. } 
\label{fig:fluka}
\end{figure}

The $\fem$ p.d.f.\ has an
energy-dependent mean (near 0.5 at 100--150~GeV,  approaching unity as $E\to\infty$).  
Its standard deviation is about 11\%, depending weakly on energy and calorimeter
composition, and it is skewed to the large-$\fem$ side~\cite{wigmansbook,groom07}.  
The mean, fractional standard deviation, and standard deviation of $\fem$ as simulated by 
FLUKA90\footnote{Since FLUKA90 many improvements  have been made, 
especially in the nuclear physics modeling.  The high-energy cascade modeling is nearly the same.  
Only the $\pi^0$ energy fraction in the cascade is of interest here.}
for cascades 
in a large lead cylinder are shown in Fig.~\ref{fig:fluka}.  For this case I obtain the empirical fits
\begin{eqnarray}
\vev{\fem} &=& 1 - \left(\frac{E}{0.76\ \hbox{GeV}}\right)^{0.87-1} \label{eqn:lead1}\\
\sigma &=& 0.125\left(\frac{E}{100\ \hbox{GeV}}\right)^{-0.076}\label{eqn:lead2} \ .
\end{eqnarray}
(The functional form of Eq.~\ref{eqn:lead1} satisfies the requirement $\vev{\fem} \to 1$ as $E \to \infty$, and
$\sigma$ must slowly approach 0 as $E \to \infty$.)
Although I use these expressions for the present calculations, 
the numerical constants will be somewhat different for different
calorimeter structures. The fractional standard deviation found for a copper/quartz-fiber 
calorimeter (Fig.~4.46 in Ref.~\cite{wigmansbook}) is shown in Fig.~\ref{fig:fluka}  for comparison.

\begin{figure}
\centerline{\includegraphics[width=3in]{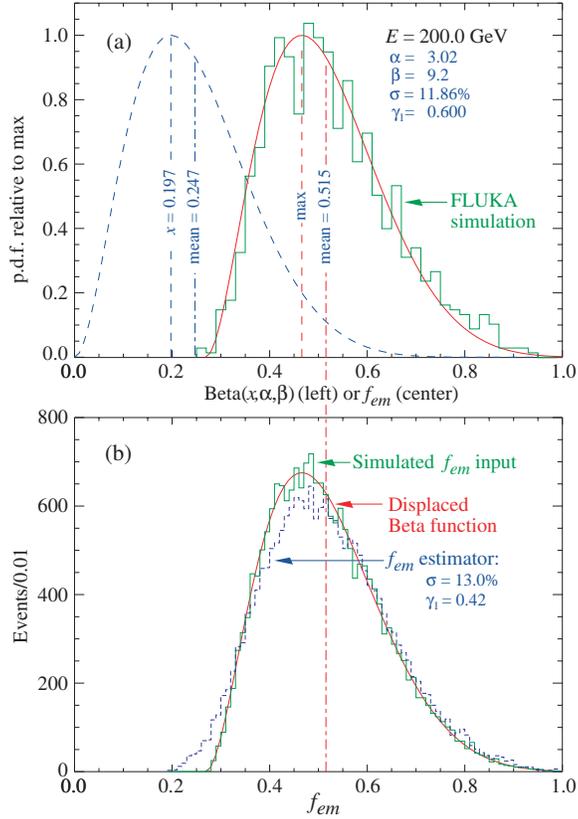}}
\caption{Approximation of the $\fem$ p.d.f.\ by a displaced Beta p.d.f.\ for 200~GeV pions incident
on a lead ``calorimeter.''
(a) The Beta p.d.f.\ (dashed curve) with $\sigma$ from Eq.~\ref{eqn:lead2} and 
$\gamma_1=0.6$, and the distribution displaced to the mean given by Eq.~\ref{eqn:lead1} 
(solid curve).  The superimposed histogram is from a FLUKA90 simulation. 
(b) A histogram of 20\,000 MC values chosen from this 
p.d.f., compared with the displaced Beta function.  The dashed histogram shows the
corresponding  $\fem$ estimator distribution obtained via
Eq.~\ref{femSolution} for  the case $\vev{\eta_C} = 0.25$ and $\vev{\eta_S} = 0.50$.  It is broader and
less skewed than the input distribution.}
\label{fig:betadist}
\end{figure}

Values of $\fem$ must be chosen from distributions with the means and r.m.s.\ widths given by 
Eqs.~\ref{eqn:lead1} and~\ref{eqn:lead2}, and with skewness that agrees with the detailed hadronic cascade
simulations.  This dimensionless skewness $\gamma_1 \ (= \mu_3/\sigma^3$, where $\mu_3$ is the third moment about the mean) is not well-determined from our simulations, but it is about 0.6, which I assume here.  Any smooth positive 
function bounded between 0 and 1 that can be adjusted to have these properties would be satisfactory.  I have 
found it convenient to use the Beta function
\begin{equation}
B(x;\alpha,\beta)\propto x^{\alpha-1}(1-x)^{\beta -1} \ ,
\end{equation}
with $\alpha$ and $\beta$ adjusted to obtain the desired $\sigma$ and $\gamma_1$.  It is then displaced
so that its mean is $\vev{\fem}$.  The displaced function does not quite go to zero at $x=1$, 
but it is sufficiently close to zero that the problem can be ignored.  
The procedure is illustrated in Fig.~\ref{fig:betadist}(a) for $E=200$~GeV pions in lead.  
The distribution from the FLUKA simulation is superimposed.

Values of $\fem$ are chosen from this function by choosing uniformly distributed random 
points $(\fem,y)$ and retaining the values of $\fem$ where  $y$ is not above the p.d.f.\
at that value of $\fem$.  A histogram of one such array of values is overlaid on the model 
p.d.f.\ in Fig.~\ref{fig:betadist}(b).  

Experimental distributions of $\fem$ obtained with the DREAM detector are shown in 
Refs.~\cite{Wigmans_Perugia_04}, \cite{DREAM05}, and \cite{tucson97}.  Because of resolution effects they are
considerably broader and less skewed than the FLUKA-generated  distributions.  
The typical $\fem$ estimator distribution shown in Fig.~\ref{fig:betadist}(b)
illustrates the resolution loss.


\section{Resolution contributions}\label{sec:resolutionsection}

In a dual-readout calorimeter $\fem$ has been  elevated from a stochastic 
quantity to a measured quantity.   In the absence of 
other resolution contributions, $C$ and $S$ are completely correlated, and the reconstructed
energy distribution given by  Eq.~(\ref{eqn:Elongform}) or Eq.~(\ref{eqn:Eshortform}) is
a delta function.
 
The signal distribution is broadened by photoelectron (p.e.) statistics, uncertain shower leakage
corrections, uncorrected signal collection irregularities, electronic noise, sampling 
fluctuations (in the case of a sampling calorimeter), intrinsic  fluctuations in the visible fraction 
of hadronic energy deposition, and other effects.  As a matter of convenience in this discussion, 
I ignore the contributions from the readout-associated factors or assume that their contributions 
are lumped into $\sigma^{\rm p.e.}$.   Although published resolution measurements  
often involve constants and other deviations from $1/\sqrt{E}$ scaling, for our 
present purposes I use
\begin{equation} 
 \frac{\sigma_E}{E} = \frac{ \sigma^{\rm p.e.} }{ \sqrt{E} }\oplus 
    \frac{\sigma^{\rm intr}} {\sqrt{E}} \oplus 
    \frac{\sigma^{\rm samp}} {\sqrt{E}}  \ ,
    \label{eqn:resolution}
 \end{equation}
where the $\sigma$'s on the right side are fractional resolutions at 1~GeV if $E$ is in~GeV.  The
hadronic intrinsic and sampling contributions ($\sigma^{\rm intr}$ and $\sigma^{\rm samp}$) are 
discussed below.

The nuclear dissociation energy deposit distribution is distinctly non-Gaussian; for example, see Fig.~4 in Ref.~\cite{tucson97}.  But visible energy deposit via ionization by charged
particles is more important in materials of interest, and its distribution is not completely correlated with that of 
nuclear dissociation and other missing energy.  The resulting visible energy deposit distribution in 
compensated sampling calorimeters seems
to be near-Gaussian, and Drews et al.~\cite{drews90} assume Gaussian distributions in separating 
intrinsic and sampling contributions.  Gaussian distributions are  assumed in Eq.~\ref{eqn:resolution} 
and elsewhere in this analysis.

 \vskip0.1in
 \noindent 1. {\em Readout statistics.}  Photomultipliers or 
 avalanche photodiodes will likely be used to detect the scintillation and Cherenkov light.  
 In this case Gaussian  (Poisson) fluctuations with variance equal to 
the number of detected photons, and hence proportional to the energy, are added to 
the signals. The fractional uncertainty thus scales as $1/\sqrt{E}$. 
  
 Electromagnetic calorimeters provide guidance about obtainable resolution.  
 Some of these are tabulated in Sec.~28.9.1  of Ref.~\cite{RPP12}. 
 Although fractional resolution as good as $2\%/\sqrt{E}$ has been obtained in 
 scintillating  crystal calorimeters, the best reported Cherenkov response in lead-glass EM
 calorimeters was 5\%, for the OPAL endcap~\cite{OPAL91}, 
 corresponding to 400~p.e.'s/GeV.   To maximize the number of photoelectrons 
 collected, $\approx\,$45\% of the ends of the lead-glass blocks were covered by the 
PMT's. Such collection efficiency will probably not be possible with the proposed crystal calorimeter.

At the other extreme, the DREAM detector obtained $35\%/\sqrt{E}$ (8~p.e./GeV) 
with quartz fibers~\cite{DREAM04,DREAM05b,DREAM05d}.  Improving Cherenkov light yield 
has been a major goal of the collaboration's recent work with a variety of 
crystals~\cite{Ak05,Ak08,Ak08bb,Ak09,Ak09bb,Ak09cc,dopedPbWO4,Ak11}.  Yields as high as
55~p.e./GeV have been reported for Mo-doped PbWO$_4$~\cite{dopedPbWO4}. 
Further improvements are possible, for example by extending the UV response of the photodetectors and 
UV transmission of the radiators.

In this model study it is realistic to assume a middle ground, 
$\sigma^{\rm p.e.}_C = \sigma^{\rm p.e.}_{0,C} /\sqrt{C}$, where $\sigma^{\rm p.e.}_{0,C}
\approx 10$--20\%, or 100--25~p.e.'s/GeV.  In most of the MC's reported here, 15\%, or 44~p.e.'s/GeV was used.

In most of the dual-readout homogeneous calorimeters under discussion, the Cherenkov and 
scintillator signals are to be separated by color and pulse shape.  This requires that the 
scintillation signal is not large compared with the Cherenkov signal, 
say $\vev{S}  = \xi^2 \vev{C}$,  where $\xi^2$ is ``a few.''   We take $\xi^2=4$ in this study:
$\sigma^{\rm p.e.}_{0,S} = \sigma^{\rm p.e.}_{0,C}/\xi$.

 \vskip0.1in

\noindent 2. {\em Sampling and intrinsic fluctuations}.  
 In a sampling calorimeter most of the energy is deposited in the absorber, 
 and there are large fluctuations in the small  fraction of the visible hadronic energy 
 deposited in the active medium.  A homogeneous calorimeter is not subject to these sampling 
fluctuations; in fact, this is an important  reason for choosing it.%
\footnote{Some suggested schemes, such 
as alternating lead glass Cherenkov planes with heavy glass scintillator 
planes~\cite{zhao06}, are totally active but only quasi-homogeneous.  In these cases sampling fluctuations have some importance.}

Much  of the hadronic energy deposit is invisible, going to nuclear disassociation, the 
production of unseen neutrons, etc., with consequent ``intrinsic'' fluctuations in the 
visible signal even if there is no absorber.  In all but a few dedicated 
test-beam experiments, sampling and intrinsic fluctuations are inextricable.  
Drews et al.~\cite{drews90} studied the problem using  compensated sandwich calorimeters with 
 separate readouts  for odd- and even- numbered layers. Since sampling
fluctuations from layer to layer are independent,  the \it sum \rm of the 
odd- and even-layer signals, and their \it difference\rm\ have the same variance. 
Signals from the intrinsic fluctuations are correlated between layers, so 
sums and differences could be used to separate sampling and intrinsic variances. 
 ($\sigma^{\rm p.e.} \approx7\%/\sqrt{E}$ did not significantly broaden the responses.) 
 In the case of lead plates, the sampling contribution was $(41.2\pm0.9)\%/\sqrt{E}$ 
 and the intrinsic contribution was $(13.4\pm4.7)\%/\sqrt{E}$.  
 In the case of uranium plates, the sampling contribution was 
 $(31.1\pm0.9)\%/\sqrt{E}$ and the intrinsic contribution was $(20.4\pm2.4)\%/\sqrt{E}$. 
Lead is probably closest to   compositions likely to be considered for the
homogeneous calorimeter, so this result is relevant to the present discussion.

\begin{table}
 \caption{Examples of near-compensating sampling hadron calorimeters.  For our present
purposes some calorimeter structure variation and constant terms in the fitted resolution have been ignored. }
\begin{center}
\begin{tabular}{ccc ccc}
\hline
\hline
Calorimeter  & Passive & Active & Resolution & $h/e$ & Reference\\
\hline
(Akesson et al.)
                      & Cu, U/Cu, U &Scint (2.5mm)& 36\%/\hbox{$\sqrt{E}$}   &$0.90$&
                      \protect\cite{UCuscint_Ake85} \\
HELIOS
                      & U (3 mm) & Scint (2.5 mm) & 34\%/\hbox{$\sqrt{E}$}   &$0.984\pm0.006$&
                      \protect\cite{HELIOScompensation_Ake87} \\
ZEUS FCAL
        &U (3.0/3.2 mm) & Scint (2,5/3.0 mm)& 35\%/\hbox{$\sqrt{E}$} & 1.03 & 
        \cite{ZEUScomp_And91,ZEUSbluebook}\\
WA80  & U (3 mm) & Scint (3 mm) &  67\%/\hbox{$\sqrt{E}$}   & 
0.89&\cite{WA80comp_You89}\\
\it (Drews, et al.)  &\it  Pb (10 mm) & \it Scint (2.5 mm) &
           \it  44\%/$\hbox{$\sqrt{E}$}$   &$ \it 0.90\pm0.01$&
\cite{drews90,Bern87}\\
(Drews, et al.)  & U (3.2 mm) & Scint (3.0 mm) &  36\%/$\hbox{$\sqrt{E}$}$   &$0.99\pm0.01$&
\cite{drews90}\\
SPACAL & Pb ($4\times$ scint vol) & 1 mm scint fibers  &  30\%/\hbox{$\sqrt{E}$}   & 0.87&
\cite{spacal_Aco91c,spacal_Arm98}\\
PCAL$^*$  & Pb$^\dagger$(10 mm) & Scint (3 mm) &  32\%/\hbox{$\sqrt{E}$}  & 
0.89&\cite{PCAL92}\\
\hline
\hline
\multicolumn{6}{l}{\footnotesize $*$ $E \le 6.8$ GeV.} \\
\multicolumn{6}{l} {\footnotesize $\dagger$ Every 6th plate is 16 mm thick Fe.}\\
\label{tab:comp_table}
\end{tabular}
\end{center}
\end{table}


Since $\fem$ fluctuations do not contribute in a compensating  calorimeter (by definition of 
``compensating''), its resolution is the result of p.e., sampling,  and intrinsic variations.%
\footnote{This is not quite true.  Suppose, for example, that $\vev{h/e} =1$, and we select 
a subset of events for which $h/e$ fluctuates to 10\% below the mean, to 0.9.  
For these events $S$ is sensitive to fluctuations of $\fem$.  The \it distribution\rm\ of 
$h/e$ about unity thus introduces sensitivity to  $\fem$.}
Examples of near-compensating calorimeter resolution are 
shown in Table~\ref{tab:comp_table}. We may take $\sigma_E=35\%/\sqrt{E}$ as a representative ``best case.''  Drew et al.'s Pb-plate calorimeter achieved
$44\%/\sqrt{E}$ (italicized line in Table~\ref{tab:comp_table}).  Scaling to this resolution, 
$\sigma^{\rm intr.}_S\approx(35/44)\times13.4\%/\sqrt{E}\approx11  \%/\sqrt{E}$.  We adopt this as the fiducial scintillator intrinsic resolution for this study.
If $\vev{\eta_S} \ne 1$, then $\sigma^{\rm intr.}_S =0.11\vev{\eta_S}/\sqrt{E}$.

The standard deviation $\sigma^{\rm intr.}_C $ of $\eta_C$ is more problematic.  
Light produced by relativistic
pions, with smaller contributions from other hadrons and electrons produced via nuclear 
$\gamma$-ray interactions, contribute a hadronic component to the Cherenkov response.  
The QFCAL~\cite{qcal97} and DREAM~\cite{DREAM05} collaborations found 
$\vev{\eta_C} = 0.20\hbox{--}0.25$. Since most
of the scintillation signal is produced by nonrelativistic ionizing particles,
$\eta_S$ and $\eta_C$ are nearly independent. 

Typically about  35 (relativistic) $\pi^\pm$'s are produced by a 100~GeV pion showering in 
Pb.\footnote{Section 2.3.1.3 of Ref.~\cite{wigmansbook}.}
The fractional standard deviation is  $1/\sqrt{35} = 0.17$ at 100~GeV, or $1.7/\sqrt{E}$.
This broad distribution is scaled
down by $\vev{\eta_C} = 0.20$--0.25:  $\sigma^{\rm intr.}_C \approx  1.7\vev{\eta_C}/\sqrt{E}$.  

\begin{figure}
\centerline{\includegraphics[width=4in]{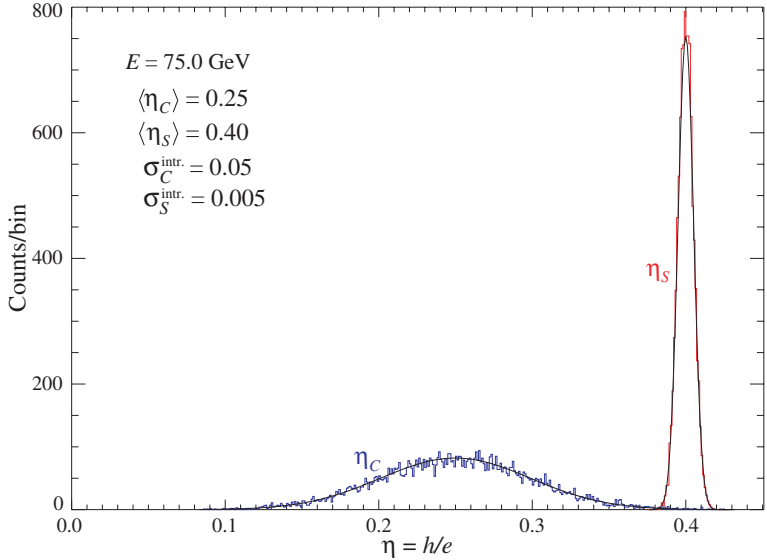}}
\caption{Input p.d.f.'s and MC realizations of the $h/e$ distributions at 75~GeV
for $\vev{\eta_C}=0.25$ and $\vev{\eta_S}=0.40$. $\sigma^{\rm intr.}_S$ and 
$\sigma^{\rm intr.}_C$ are from Table~\ref{tab:parameters}.
}
\label{fig:input_intrinsic}
\end{figure}

These results are summarized in Table~\ref{tab:parameters}.  Typical distributions 
of $\eta_S$ and $\eta_C$ are shown in Fig.~\ref{fig:input_intrinsic}.
Since the energy distributions reported for compensating calorimeter, e.g., as given in 
Table~\ref{tab:comp_table} are all consistent with Gaussian distribution, 
it is evidently valid to consider the
distributions discussed in this section as Gaussian as well. This results in near-Gaussian 
energy-estimator distributions as reconstructed from $S$ and $C$.
 
\begin{table}
\caption{Summary of resolution contributions.  Reference values in the last column
are used in Sec.~\ref{sec:reconstructed}.}
\begin{center}
 \begin{tabular}{llll}
  \hline
Std dev & Value at $E$ & Reference value&Optimistic value\cr
 \hline
  $\sigma^{\rm p.e.}_C$ & $\sigma^{\rm p.e.}_{0,C}/\sqrt{C}$ & $\sigma^{\rm p.e.}_{0,C}=15\%$
   &$\sigma^{\rm p.e.}_{0,C}=5\%$\cr
$\sigma^{\rm p.e.}_S$ & $\sigma^{\rm p.e.}_{0,C}/\xi\sqrt{S}$ & $\xi=2$ & $\xi=10$ \cr
$\sigma^{\rm intr.}_C$& $\sigma^{\rm intr.}_{0,C} \vev{\eta_C}/\sqrt{E}$ & 
$\sigma^{\rm intr.}_{0,C}=170\%$ &$\sigma^{\rm intr.}_{0,C}=85\%$\cr
$\sigma^{\rm intr.}_S$ & $\sigma^{\rm intr.}_{0,S}\vev{\eta_S}/\sqrt{E} $ &
 $\sigma^{\rm intr.}_{0,S}=11\%$ &5.5\%\cr
\hline
\label{tab:parameters}
 \end{tabular}
\end{center}
 \end{table}



\section{Simulation of the energy estimator distribution}\label{sec:reconstructed}

Equation \ref{eqn:Eshortform} and the equivalent Eq.~\ref{eqn:Elongform} describe the 
estimator of  incident pion energy as obtained from the observed scintillation
and Cherenkov  signals.   We incorporate 
the resolution contributions discussed in Sec.~\ref{sec:resolutionsection} and summarized in 
Table~\ref{tab:parameters} to obtain distributions of $S$, $C$, and the energy estimator $E$
via  a simple, transparent Monte Carlo calculation as follows:

\vskip0.1in\begin{enumerate}
\item Choose the incident energy $E$ and the detection efficiency ratios $\vev{\eta_S}$ and 
$\vev{\eta_C}$, fixing the latter in the range 0.20--0.25.

\item Choose the resolution parameters, using the reference values given in 
Table~\ref{tab:parameters}.

\item Generate an array of $N$  values of $\fem$ chosen from the 
displaced Beta distribution for energy $E$.
\item Generate $N$ values of  $\eta_S$ from a normal distribution with mean 
$\vev{\eta_S}$ and fractional standard deviation $\sigma^{\rm intr.}_S$.  
  Similarly, generate $N$ values of $\eta_C$ from a normal
distribution  with mean $\vev{\eta_C}$ and standard deviation $\sigma^{\rm intr.}_C$. 

\item Construct the corresponding  $S$ and $C$ arrays via Eqs.~\ref{eqn:S} and~\ref{eqn:C}.

\item Replace each $S$ and $C$ as calculated in step 4 with values chosen from
normal distributions with means $S$ and $C$ and standard deviations 
$\sigma^{\rm p.e.}_{0,C}\sqrt{C}$ and $\sigma^{\rm p.e.}_{0,S}\sqrt{S} $, respectively. 
The resulting  $S$ and $C$ include p.e.\ statistics, and are used for  subsequent ``data analysis.''
(Since $\sigma^{\rm p.e.}_{0,C}/\sqrt{C}$ and $\sigma^{\rm p.e.}_{0,S}/\sqrt{S}$
are fractional standard deviations,
$\sigma^{\rm p.e.}_{0,C}\sqrt{C}$ and  $\sigma^{\rm p.e.}_{0,S}\sqrt{S}$ have the same units 
as $S$ and $C$ (GeV).)

\item Find the energy estimator array via Eq.~\ref{eqn:Elongform} or Eq.~\ref{eqn:Eshortform}.

\end{enumerate}

\begin{figure}
\centerline{\includegraphics[width=6in]{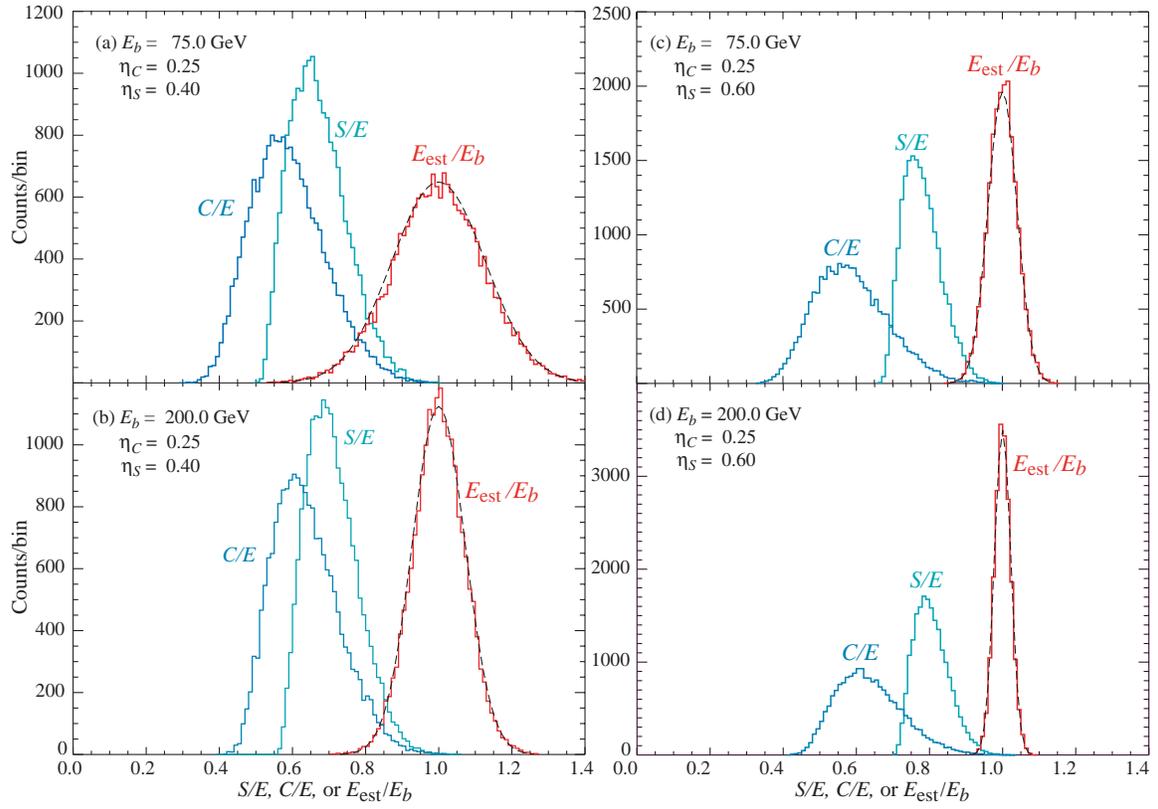}}
	\caption{Monte Carlo distributions of $C$, $S$, and the estimator of $E$.  (a) and (c) are
	 for beam energies of 75~GeV; for (b) and (d) the beam energy is 200~GeV. 
	 For (a) and (b) $\eta_S = 0.40$, while for (c) and (d) 
	$\eta_S = 0.60$.  In all cases $\eta_C = 0.25$.   Gaussians with the ``measured'' 
	$\sigma_E$ and mean relative to beam energy are shown as dotted black lines.  Details
	are given in Table~\ref{tab:examples_table}.}
	\label{fig:fourpanel}
\end{figure}

The results of four 10\,000 event simulations 
are shown in Fig.~\ref{fig:fourpanel}, at 75 and 200 GeV using  realistic (0.4) and 
optimistic (0.6) values of $\eta_S$.  In each case
the mean value of the estimator of $E\ (E_{\rm est})$ scaled by the beam energy is 1.00.   
The distributions agree well with the Gaussians  with the same mean and 
standard deviation  drawn over the $E$ histograms. In nearly 
all cases $|\gamma_1| \leq 0.05$.

The fractional standard deviation of the energy estimator scales as $1/\sqrt{E}$.  The coefficient (resolution at 1~GeV) as a function of $h/e|_S\ (\equiv\eta_S)$ is shown in Fig.~\ref{fig:resvhoeS} for two values of 
$h/e|_C\ (\equiv\eta_C)$ that bracket the range found for quartz-fiber 
readout calorimeters~\cite{qcal97,Wigmans_Perugia_04,DREAM05}.

\begin{table}
 \caption{Parameters for the  examples shown in Fig.~\ref{fig:fourpanel}, each based on 20\,000 simulated events.  In all cases $\eta_C = 0.25$.}
\begin{center}
\begin{tabular}{ccc cc}
\hline
\hline
  & (a) & (b) & (c) & (d)\\
 \hline
  $E_b$ & 75 GeV  & 200 GeV & 75 GeV & 200 GeV \\
$\eta_S$&0.40 & 0.40 & 0.60& 0.60\\
 \hline
 $\vev{S/E_b}$     & 0.67      & 0.71     & 0.78      & 0.81 \\
 $\sigma_S$         & 11.6\% & 10.1\% & 6.7\% & 5.9\% \\
 $\gamma_{1S}$ & 0.57     & 0.62      & 0.58    & 0.59 \\
 $\vev{C/E_b}$     &0.59        & 0.64       & 0.59           & 0.64 \\
 $\sigma_C$         & 17.2\% & 14.2\%  & 17.2\%    & 14.2\%  \\
 $\gamma_{1C}$ & 0.42        & 0.55       & 0.45           & 0.53 \\
$\vev{E_{\rm est}/E_b}$ & 0.999 & 1.000 & 1.000 & 1.000\\
 $\sigma_E$ &  12.8\% &7.24\% & 3.96\% & 2.27\%\\
 $\sigma_E$ @ 1 GeV & 111.\%  & 102.\% & 34.3\% &32.1\% \\
$\gamma_{1E}$ & $0.008$ & $0.002$ & $-0.015$ & $-0.003$ \\
 \hline
\hline
\label{tab:examples_table}
\end{tabular}
\end{center}
 \vskip-0.1in
\end{table}


\section{Discussion and conclusions}

The standard deviations described in this section are highly uncertain, and can be used only as guides.  The effect of large changes in the input variances on the resolution curves shown in Fig.~\ref{fig:resvhoeS} have been studied.   Examples of optimistic
excursions from best estimates are given in Table~\ref{tab:parameters}.  
At $\vev{\eta_S}=0.45$ and $\vev{\eta_C} = 0.20$, where $\sigma_E=52.5\%/\sqrt{E}$, the following improvements are found:
  
  \begin{figure}
\centerline{\includegraphics[width=4in]{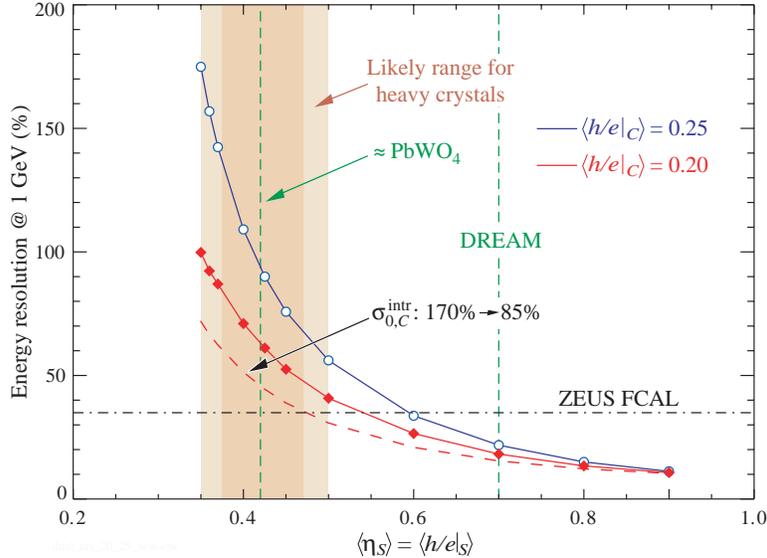}}
\caption{Dependence of resolution on $\eta_S$ at two values of $\eta_C\equiv h/e|_C$.
The results for $\eta_C=0.20$ are essentially those for $\eta_C=0.25$ displaced to the
left by 0.05; it is the \it contrast\rm\ between $\eta_C$ and $\eta_S$ that determines
the resolution.   The dashed line shows the effect of halving the width of the $\eta_C$ p.d.f.}
\label{fig:resvhoeS}
\end{figure}

\begin{enumerate}

\item Increase the Cherenkov p.e.~yield by a factor of nine, from 
$\sigma^{\rm p.e.}_{0,C}=15\%$ (44 p.e.'s/GeV) to 
$\sigma^{\rm p.e.}_{0,C}=5\%$ (400 p.e.'s/GeV).  The resolution improves from
$52.5\%/\sqrt{E}$ to $47.1\%/\sqrt{E}$, or $-$5.4\%---not a large improvement.

\item Increase the scintillator p.e.~yield by a factor of 25 ($\xi = 2 \to \xi=10$). The resolution 
improves from $52.5\%/\sqrt{E}$ to $49.0\%/\sqrt{E}$, or $-$3.5\%.  
At least in this model, there is no serious 
penalty for using a ``weak'' scintillator.

\item Decrease the width of the Cherenkov  intrinsic hadronic resolution at 1~GeV
by a factor of two, from 170\% to 85\%.  The former number is expected from fluctuations 
in the number of relativistic pions produced in the cascade, and so
should be relatively dependable. However, this change produces a large
improvement:  $52.5\%/\sqrt{E} \to 39.1\%/\sqrt{E}$, or $-$13.4\%.  
The curve for this case has been added to Fig.~\ref{fig:resvhoeS}.

\item Finally, halve the width of the scintillator intrinsic hadronic resolution 
from $11\%/\sqrt{E}$ to $5.5\%/\sqrt{E}$. As expected, there is
little effect: $52.5\%/\sqrt{E} \to 52.1\%$, or $-$0.4\%. A substantial 
downward excursion of this number would be unphysical.

\end{enumerate}

The overall conclusion remains and is evident from 
Eq.~\ref{eqn:Elongform}:  in a homogeneous hadronic calorimeter with dual readouts and 
a non-hydrogenous scintillator,  event-to-event fluctuations of $\eta_C$ and $\eta_S$ 
destroy the resolution if their means are close.  A large contrast between them is unlikely
because the mechanisms that increase $\vev{\eta_S}$ are not present.
It remains true that the mean of the energy estimator distribution is the beam energy, and its 
distribution is nearly Gaussian.

Detailed simulations with a modern, sophisticated Monte Carlo program such as GEANT4 would be desirable before commitment to a large R\&D effort. It would be particularly informative to understand the means and variances of $\eta_C$ and $\eta_S$.

\section*{Acknowledgments}

I particularly thank Adam Para for getting me interested in this problem.  Richard Wigmans'
book, many papers, and many personal interactions have been invaluable, as have been discussions with John Hauptman.  The understanding  of hadronic calorimetry 
underlying this paper has been slowly forged over at least the past 40 years 
by the insights and  hard work of experts too numerous to mention.

This work was supported by the U.S. Department of  Energy under Contract No.\ DE-AC02-05CH11231.

\end{document}